\documentclass{acm_proc_article-sp}
\usepackage{algorithm}
\usepackage{algorithmic}

\usepackage{amssymb}
\usepackage{amsmath}
\usepackage{subfigure}
\usepackage{graphicx}
\usepackage{enumerate}
\usepackage{verbatim}

\begin{document}
\title{A Navigation Algorithm Inspired by Human Navigation}
\numberofauthors{5} 
\author{
\alignauthor
Vijesh M \\ \affaddr{intern\\Indian Statistical Institute}\\
       \affaddr{Chennai, India}\\
       \email{mv.vijesh@gmail.com}
\alignauthor
Sudarshan Iyengar\\ \affaddr{Indian Statistical Institute}\\
       \affaddr{Chennai, India}\\
       \email{sudarshaniisc@gmail.com}
\alignauthor Vijay Mahantesh SM\\
       \affaddr{Intern \\Indian Statistical Institute}\\
       \affaddr{Chennai, India}\\
       \email{vijaym123@gmail.com}
\and  
\alignauthor Amitash Ramesh \\
       \affaddr{Intern}\\
       \affaddr{Indian Statistical Institute}\\
       \affaddr{Chennai, India}\\
       \email{amitashr@gmail.com}
\alignauthor Veni Madhavan \\
       \affaddr{Computer Science and Automation}\\
       \affaddr{Indian Institute of Science}\\
       \affaddr{Bangalore, India}
}
\maketitle
\begin{abstract}

Human navigation has been a topic of interest in spatial cognition 
from the past few decades. It has been experimentally observed 
that humans accomplish the task of way-finding a destination in an 
unknown environment by recognizing landmarks. Investigations using 
network analytic techniques reveal that humans, when asked to 
way-find their destination, learn the top ranked nodes of a 
network.  In this paper we report a study simulating the strategy 
used by humans to recognize the centers of a network. We show that 
the paths obtained from our simulation has the same properties as 
the paths obtained in human based experiment. The simulation thus 
performed leads to a novel way of path-finding in a network. We 
discuss the performance of our method and compare it with the 
existing techniques to find a path between a pair of nodes in a 
network. 
\end{abstract}

\section{Introduction}

The problem of finding a target vertex in a network where one is provided 
with just the local information is a well studied problem. Starting from 
the work of Milgram in 1967~\cite{milgram67}, the current methods include 
routing using full-tables~\cite{gavoille01}, interval 
routing~\cite{gavoille99,gavoille01}, routing labeling 
schemes~\cite{peleg00,thorup01}, greedy routing~\cite{giordano01,kleinberg-1-00}, 
geographic routing~\cite{giordano01}, compass routing~\cite{giordano01}, 
etc., These routing techniques are mainly used in transportation networks 
and in wired as well as wireless communication networks. For detailed 
survey one can refer to~\cite{gavoille99, gavoille01, giordano01} 
and~\cite{peleg00}. 

For surveys of the routing methods in social networks and email networks 
in particular, one is referred to~\cite{adamic02, adamic01, fraigniaud07, kleinberg-1-00, nowell05}

%
 
In the classical small world experiment by Milgram~\cite{milgram67},
the letter from Nebraska was to reach Boston. This is a scenario where 
the sender actively needs to participate in the process of helping the 
letter reach the destination, while the receiver remains passive. Let 
us  consider a variation of this problem where both sender and 
receiver participate in the process of finding a path between them. 

{\bf Motivation}

Sudarshan et al., in \cite{sudarshan11} presents a network analytic approach
to understand human navigation. Their experiment comprises of making human
participants navigate on a network that is unknown to them. Participants
are given a source node and are asked to navigate to a destination node. 
The underlying network is not visible to the participant in its entirety. 
At any instance, participant can view the adjacent nodes of the node that
s/he is currently present.  Several such source destination node pairs are
given to the participant. The time taken and the path taken are recorded.

They observe that participants eventually learn a strategy which 
helps them navigate easily and quickly. They show that the participants
learn a few landmark nodes which they use as via media to navigate. They further
 show that these landmark nodes have superior closeness-centrality
ranking. 

In the current paper, we present a \emph{navigation algorithm} based on the strategy
used by human participants in the experiment reported in \cite{sudarshan11}.

The strategy of human participants in the above mentioned experiment to learn 
centers of a network and the pursuit of paths through the centers has motivated 
the proposal of this novel navigation algorithm. 

{\bf The Technique}

In an unweighted graph $G(V,E)$, given a source vertex $s$ and a 
destination vertex $t$, a straight forward approach to establish 
a path between $s$ and $t$ is to take a random walk from $s$ till 
we reach $t$. One way to better this random walk method is to take 
a two way random walk, one from the source $s$ and the other from 
the destination $t$ and stop once the two random walks intersect. 
This method is expected to be quicker than the first in terms of 
number of hops that is required to establish a path. We provide 
empirical results to support this. 

We describe our algorithm (\emph{path concatenation 
algorithm} or PCA) and compare it with two other path finding 
strategies based on random walks. We apply PCA on various 
synthetic networks and discuss its performance. We show that PCA 
yields center-strategic paths on scale free networks - as was the 
case with the paths that the participants took in the two experiment.

We start with the description of related work in Section.~\ref{sec:4_related_work}
and then in Section.~\ref{sec:4_notions} present the notations and notions used. 
Section.~\label{4_results} comprises of the application of the described algorithm
on synthetic networks. We benchmark our results by comparing them 
with two other navigational techniques. In Section.~\ref{sec:4_conclusion} we conclude 
with a possible further improvements.

\section{Related Work}
\label{sec:4_related_work}

Kleinberg's work~\cite{kleinberg-2-00} on navigation in a small world, was the 
first paper to shed light on the navigation problem on complex networks. The
paper highlighted the fact that {it is easier to find short chains between 
points in \emph{some} networks than \emph{others}}.

Adamic et al., in~\cite{adamic01} introduced several local search strategies to 
find a path to the given destination vertex. Their strategies are limited to 
applications on networks that obey the power law. The results are shown on 
GNUTELLA peer-to-peer network which is known to obey power law. In~\cite{kim02}, 
Kim et al., numerically compares the local path finding strategies with that of 
global path finding strategies on scale free networks. 

A rigorous graph theory based approach to path finding strategy was proposed 
by Gavoille et al.,~\cite{gavoille04}. They provide a method of navigation based
on vertex labeling. They establish a labeling strategy for the vertices of a graph
in a way that allows one to compute the distance between any two vertices directly
from the labels, without using any additional information about the network.

Adamic et al., in~\cite{adamic05} addresses the question of how participants 
in a small world experiment are able to find short paths in a social network 
using only local information about their immediate contacts. They conduct their 
experiment on an email network and demonstrate by empirical data that the small 
world search strategies using a contact's position in physical space or in an 
organizational hierarchy relative to the target can effectively be used to 
locate most individuals.

A detailed survey on decentralized search algorithms on networks that exhibit 
small-world phenomena is given by Kleinberg in~\cite{kleinberg06}. This survey
also contains an exhaustive list of open problems in this area.

Ozgur et al.,~\cite{simsek08} propose a navigational technique based on a hybridized
method of using degree and homophilly of vertices that yield better results than the
currently known techniques.

Cajueiro et al., in~\cite{cajueiro09} proposed a learning framework based on a 
First-Visit Monte Carlo Algorithm. They show that the \emph{navigation difficulty}
and \emph{learning velocity} are strongly related to the network topology. 
In~\cite{cajueiro-1-09}, Cajueiro has proposed a strategy where the walker is
assumed to take optimal paths in order to minimize the cost of walking. They provide
an approach to generalize several concepts presented in the literature 
concerning random navigation and direct navigation.

Recently a novel method of navigating a network using the underlying spanning tree
was proposed by Dragan et al. in~\cite{dragan11}. The paper provides a thorough
graph theoretic treatment of the navigation problem. They reduce the navigation problem
on a graph $G$ to that of the underlying spanning tree and navigate on the latter. 

On the application front, such navigational techniques are useful in a problem of
finding paths between vertices in peer-to-peer systems.  Crespo et al.,~\cite{crespo02} 
in their work on routing indices on Peer-to-Peer systems, 
propose a novel method of query forwarding from vertices to their neighbors that are 
more likely to have answers. They provide different novel routing schemes and 
evaluate their performance.

\section{Notations and Notions} 
\label{sec:4_notions}

A graph $G=(V,E)$ is composed of a set of nodes $V$ and a set of edges 
$E\subseteq V\times V$, with $|V|=n$ and $|E|=m$. A {\it way} between 
two nodes $u$ and $v$ is any sequence of edges $(e_1, e_2, \dots, e_k)$ 
with $e_1=(u,x_1), e_2=(x_1,x_2), \dots ,e_k=(x_{k-1},v)$. A {\it path} is a way with no repeating 
nodes. The length of a way is defined as the number of edges in it. 
A \emph{shortest path} between two nodes $u$ and $v$ is a path with length $l$, such that, 
every other path between $u$ and $v$ is of length greater than or equal to $l$.
Two nodes are said to be {\it connected} if there exists a path between them. The 
{\it distance} $d(v_1, v_2)$ between any two nodes is defined as the length 
of a shortest path between them, or set to $\infty$ if there exists no path 
between them. Any maximal set of pairwise connected nodes is 
called a {\it component} of the graph. Given $v \in V$, $N_{G}(v)$ denotes the set 
of all adjacent vertices of $v$. 

Let $W_u=(u=v_0,v_1,v_2,....)$ denote a random walk starting from the vertex $u$. By 
$W_u[i]$ we mean the vertex at the $i$th step of the random walk $W_u$.
e.g. $W_u[2]=v_2$. Let $T(W_u,i)$ denote the set $T(W_u,i)=\{v_0,v_1,...,v_i\}$ 
of all vertices that are being visited by the random walk $W_u$ during the first $i$ steps.

A {\it centrality index} is a real-valued function $C: V \rightarrow \mathbb R$ 
on the nodes. 
The intuition is that the higher the value of this 
function, the more central this node is for the network. There are various 
indices; in this article we use the so-called {\it closeness centrality} 
\cite{Sabidussi66} $C_C(v)$, which is defined as the reciprocal of sum of the distances of $v$ to all other nodes $w$: 
$$C_C(v) = 1/\sum_{w\in V} d(v,w)$$
For any given graph, a centrality index can be used to define a ranking on the nodes, by sorting the nodes 
non-decreasingly by their centrality value. Ties if any, are broken arbitrarily. we denote the rank of a vertex by $R(v)$.

Given a walk from $s$ to $t$, it is easy to see that one can get a path from $s$ to $t$ 
on this walk by removing the cycles.
(Refer~\cite{grimaldi03}).By $P(W_u,i)$ we mean a path obtained by considering the 
first $i$ steps taken by the random walk $W_u$.

Through out the paper, we assume that the graph is an undirected simple graph, 
unless otherwise stated.

\section{The Algorithm}
\label{sec:idea}
The path concatenation algorithm is split into two components.

\begin{itemize}
\item The Learning Phase
\item The Navigation Phase
\end{itemize}

The learning phase deals with the preprocessing of the graph G. Once 
the preprocessing phase is complete, the navigation phase describes how 
one can find a path between a given pair of vertices. 

\subsection{Learning Phase}
\label{sec:4_learning_phase}

In the learning phase, the undirected graph $G(V,E)$ is converted to a 
directed weighted graph $G_d(V,E_d)$

Let $u,v\in_r V$ be two randomly picked vertices. Consider the random walks 
$W_u$ and $W_v$. As defined above, let $W_u=(W_u[0],W_u[1],...)$ and 
$W_v=(W_v[0],W_v[1],...)$.

Consider least $i$ such that $T(W_u,i)\cap T(W_v,i)\neq \emptyset$. 

Such an $i$ guarantees a vertex $h\in T(W_u,i)\cap T(W_v,i)$. 
Clearly, one can construct a path from $u$ 
to $h$ and another path from $v$ to $h$. Let us call these directed paths 
$P_{u,h}$ and $P_{v,h}$ respectively. Let $l(P_{u,h})$ and $l(P_{v,h})$ 
denote the respective path lengths.

We define a function on the vertex set called the \emph{Vertex Reward} 
function and a function on the edge set called the \emph{Edge Reward} 
function as follows.

\subsubsection{Vertex Reward}

Consider a function $flag :V_d\rightarrow \mathbb{Z}$, where 
$flag(v)=0, \forall v\in V_d$ to begin with. In an iteration, we 
pick a random pair of vertices $u$ and $v$, and take random walks 
$W_u$ and $W_v$. Once the walks intersect at a vertex, say $h$, we 
increment $flag(h)$ by 1, i.e. $flag(h)=flag(h)+1$. 

\subsubsection{Edge Reward}

Given the graph $G(V,E)$, at the end of the learning phase, we need 
to obtain a weighted directed graph $G(V,E_d)$, with $E_d=V\times V$.
We define a reward function $R:E_d\rightarrow \mathbb{R}$. We first 
initialize $R(u,v)=0, \forall u,v \in V_d$. For every edge $(a,b)$ 
belonging to the path $P_{u,h}$, we increment the reward of that edge by
$R(a,b)=R(a,b)+\frac{1}{l(P_{u,h})}$. Similarly for all edges $(c,d)$ 
belonging to the path $P_{v,h}$, we increment the reward of the edge 
$(c,d)$ by $R(c,d)=R(c,d)+\frac{1}{l(P_{v,h})}$.

Below is the pidgin code of the algorithm. 

\newpage
\subsubsection{Algorithm for Learning Phase}

\begin{algorithm}
\label{4_learning_algorithm}
\caption{The Learning phase}
\begin{algorithmic}[1]
\renewcommand{\algorithmicrequire}{\textbf{Input:}}
\renewcommand{\algorithmicensure}{\textbf{Output:}}

\REQUIRE $G(V,E)$, $\alpha$
\ENSURE  $G_d(V,E_d)$, $R$

\STATE $R(a,b)\longleftarrow 0, \forall a,b\in V$

\STATE $flag(v)\longleftarrow 0, \forall v\in V$

\FORALL{$ (u,v) \in V\times V$}
\STATE 
\begin{itemize}
\renewcommand{\labelitemi}{$\bullet $}
\item Take random walks $W_u=(W_u[0],W_u[1],...W_u[i])$ and $W_v=(W_v[0],W_v[1],...W_v[i])$ until $T(W_u,i)$ $\bigcap$ $T(W_v,i)$ $\neq \emptyset$. \item Pick $h \in T(W_u,i)$ $\bigcap$ $T(W_v,i)$ \\
Compute $P_{u,h}$ and $P_{v,h}$\\
\item 
$l(P_{u,h})$ $\longleftarrow$ path length of $P_{u,h}$\\
$l(P_{v,h})$ $\longleftarrow$ path length of $P_{v,h}$

\item $flag(h) \longleftarrow flag(h)+1$

\item \textbf{for} every directed edge $(a,b) \in P_{u,h} $ \textbf{do}\\
\hspace{0.3 in} $R(a,b) \longleftarrow R(a,b) + \frac{1}{l(u,h)}$

\item \textbf{for} every directed edge $(c,d) \in P_{v,h} $ \textbf{do}\\
\hspace{0.3 in} $R(c,d) \longleftarrow R(c,d) + \frac{1}{l(v,h)}$

\end{itemize} 
\ENDFOR

\STATE Let $HotSpotOrdering \longleftarrow$ list of vertices in descending order of their $flag$ values (with ties arbitrarily broken).

\STATE Let the first $\alpha$ number of vertices from $HotSpotOrdering$ constitute the $HotSpots$ set. 

\STATE Compute and store the shortest paths between all pairs of vertices in the $HotSpot$ set. Given $u,v\in HotSpots$, $HotSpotLookup(u,v)$ returns a shortest path from $u$ to $v$.

\end{algorithmic}

\end{algorithm}

Algorithm~\ref{4_learning_algorithm} is a pre-processing phase where, if a graph $G(V,E)$ is given as input,
one obtains a directed weighted graph and a table of values comprising of shortest paths between 
$\alpha$ number of top $flag$ valued vertices of the graph, i.e., all pairs shortest paths amongst the 
vertices in $HotSpots$.

We now describe how one can find a path from a given source vertex $s$
to a destination vertex $t$ by making use of the weights in the directed weighted graph thus 
obtained. Information from $HotSpots$, $HotSpotLookup$ and the edge reward function $R$ is 
utilized to find a path from a source to a destination vertex. 

\begin{figure}[htp]
\centering
\includegraphics[scale=0.13]{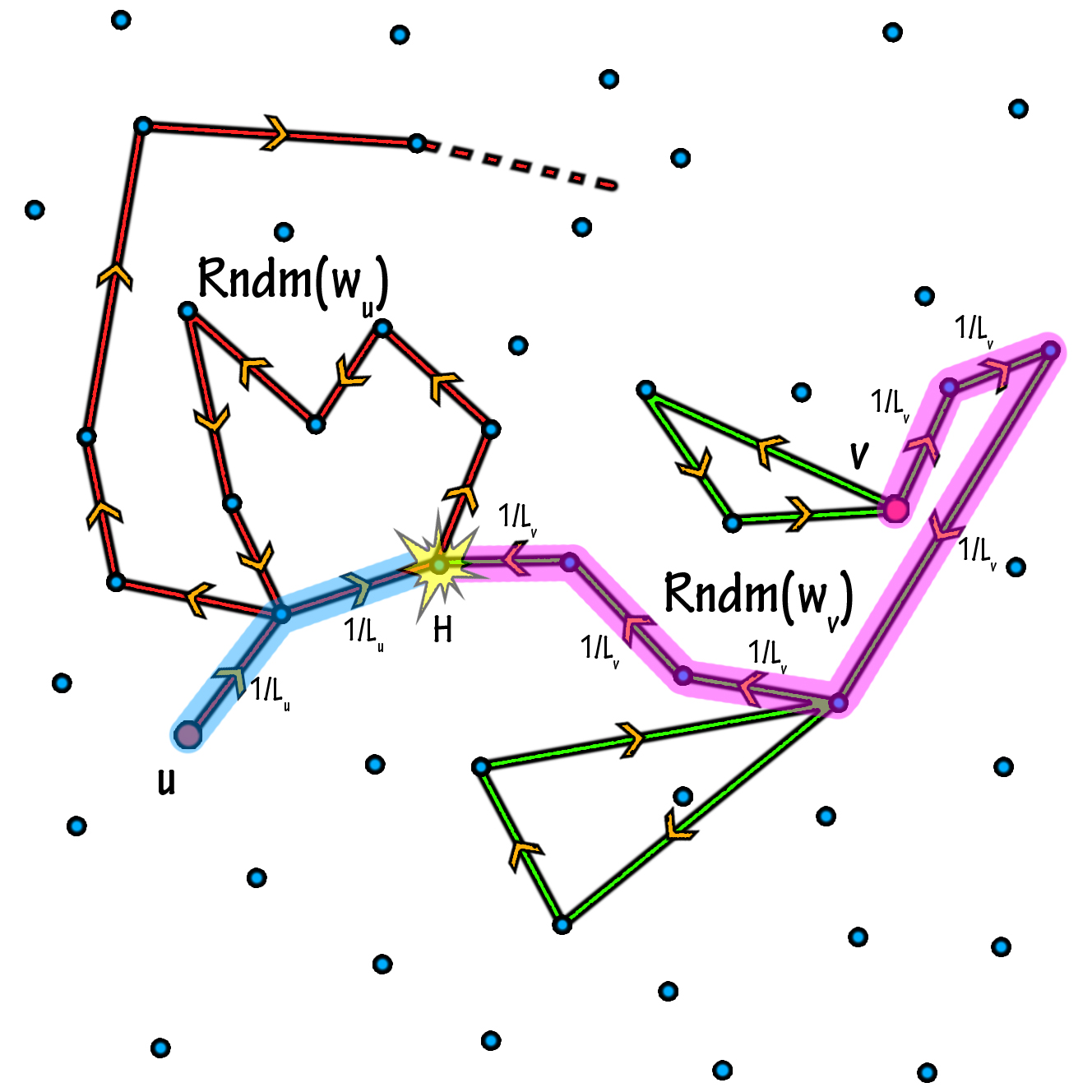}
\caption{The Learning Phase}

\label{4_learning_phase_diag}
\end{figure}

Figure.~\ref{4_learning_phase_diag} illustrates the learning phase of Algorithm \ref{4_learning_algorithm}. $u$ and $v$ are the 
chosen pair of vertices. The red walk from $u$ and the green walk from $v$ 
indicate the random walks $W_u$ and $W_v$ respectively. $h$ is the vertex at which the intersection has occurred. The blue path from 
$u$ and the pink path from $v$ indicate $P_{u,h}$ and $P_{v,h}$ respectively. $l(P_{u,h})$ and 
$l(P_{v,h})$ are the lengths of the blue and pink path respectively. The directed edges along the 
blue path and the pink path are rewarded with $\frac{1}{l(P_{u,h})}$ and $\frac{1}{l(P_{v,h})}$ 
respectively. $flag(h)$ is incremented by 1.

\subsection{Navigation Phase}
\label{sec:4_navigation_phase}

The navigation phase makes use of the output $G_d(V,E_d)$ and the set $R$, which contains values of the edges 
obtained in the learning phase to find a path between a given source vertex $s\in V$ 
to a destination vertex $t \in V$. The technique involves 3 stages:
\begin{enumerate}
\item Finding a path from $s$ to $h_s$, where $h_s\in HotSpots$
\item Finding a path from $t$ to $h_t$, where $h_t \in HotSpots$
\item Concatenating the paths $s$ to $h_s$, $h_s$ to $h_t$ and $h_t$ to $t$.
\end{enumerate}

The paths $s$ to $h_s$ and $t$ to $h_t$ are constructed by choosing adjacent edges with the highest 
$R$ value. We call this the \emph{greedy traversal technique}. This is illustrated in Figure 2. 
Assuming that we start from a start vertex $s=u$, let $u_{0}$,  $u_{1}$, $u_{2}$ $u_{3}$ and $u_{4}$ 
denote the neighbors of $u$. As can be seen in Figure 2, the edge with the highest $R$ value is $(u,u_1)$,
we choose this as the edge and continue forming a path by continuing this process ($(u_1,u_{11})$ is chosen next).

The paths thus obtained are finally concatenated to get a path from $s$ to $t$. See Figure 3 for illustration.

\begin{figure}[htp]
\centering
\includegraphics[scale=0.17]{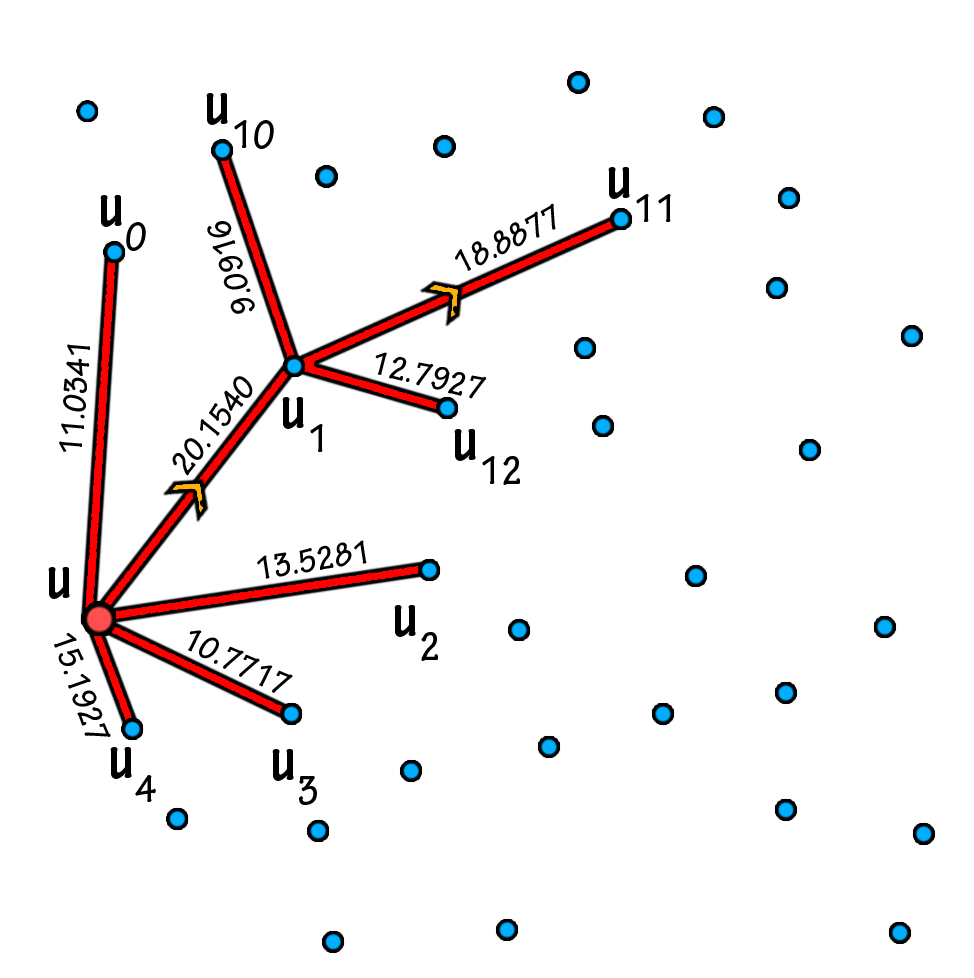}
\caption{Greedy Traversal Technique}
\label{4_greedy_traversal}
\end{figure}

\begin{figure}[htp]
\centering
\includegraphics[scale=0.15]{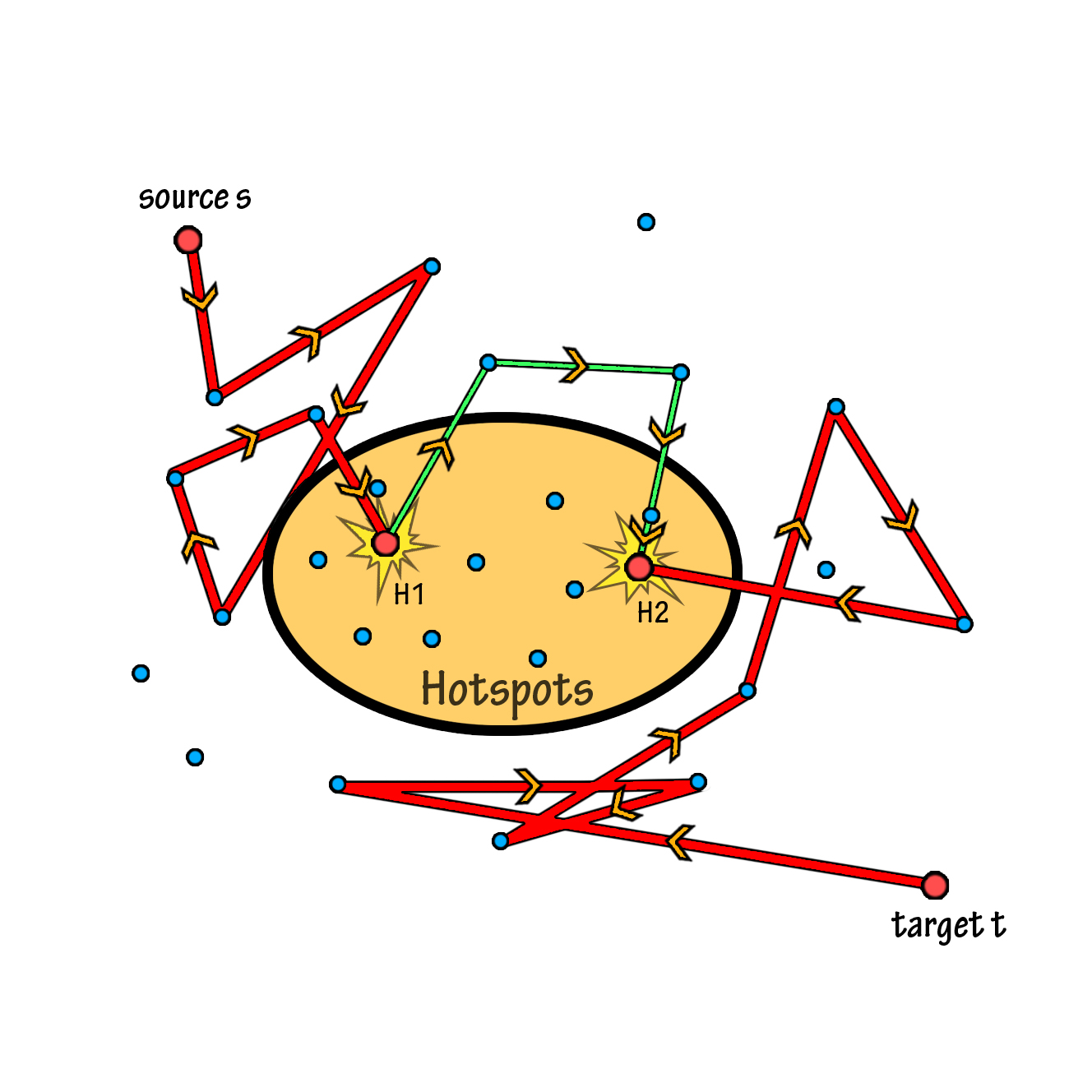}
\caption{The Navigation Phase}
\label{4_navigation_phase}
\end{figure}

\newpage
\section{\label{4_results}Results and Discussions}

We discuss in this section, the results that were obtained when the path concatenation 
algorithm was applied to several classes of graphs.

We mainly test our results on two types of synthetic networks: Barabasi-Albert (scale-free) graphs 
and Erdos-Renyi graphs.

\textit{Barabasi-Albert graphs}: Barabasi-Albert Graphs ~\cite{barabasi02} are a class of graphs 
constructed by preferential attachment of vertices.

\textit{Erdos-Renyi graphs}: Erdos-Renyi Networks ~\cite{erdos60} are a class of graphs constructed 
by including the edges with a given probability $p$.

\subsection{Selection of optimum value of $\alpha$}

\begin{figure}[h]
\centering
\includegraphics[scale=0.2]{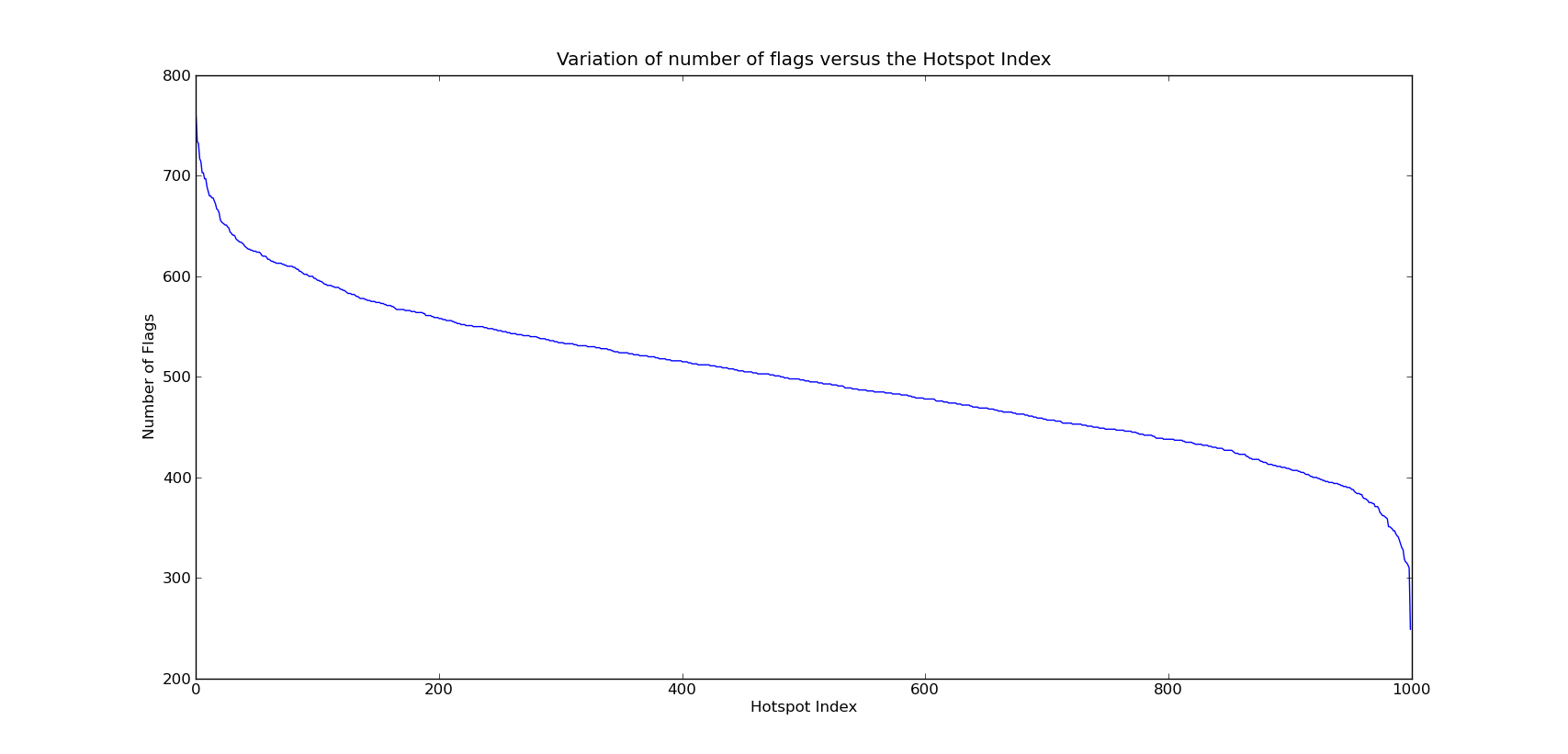}
\caption{Plot of variation of number of flags versus $HotSpotIndex$ 
for an Erdos-Renyi network of 1000 vertices
and edge probability of 0.15}
\label{4_optimum_er}
\end{figure}

\begin{figure}[h]
\centering
\includegraphics[scale=0.2]{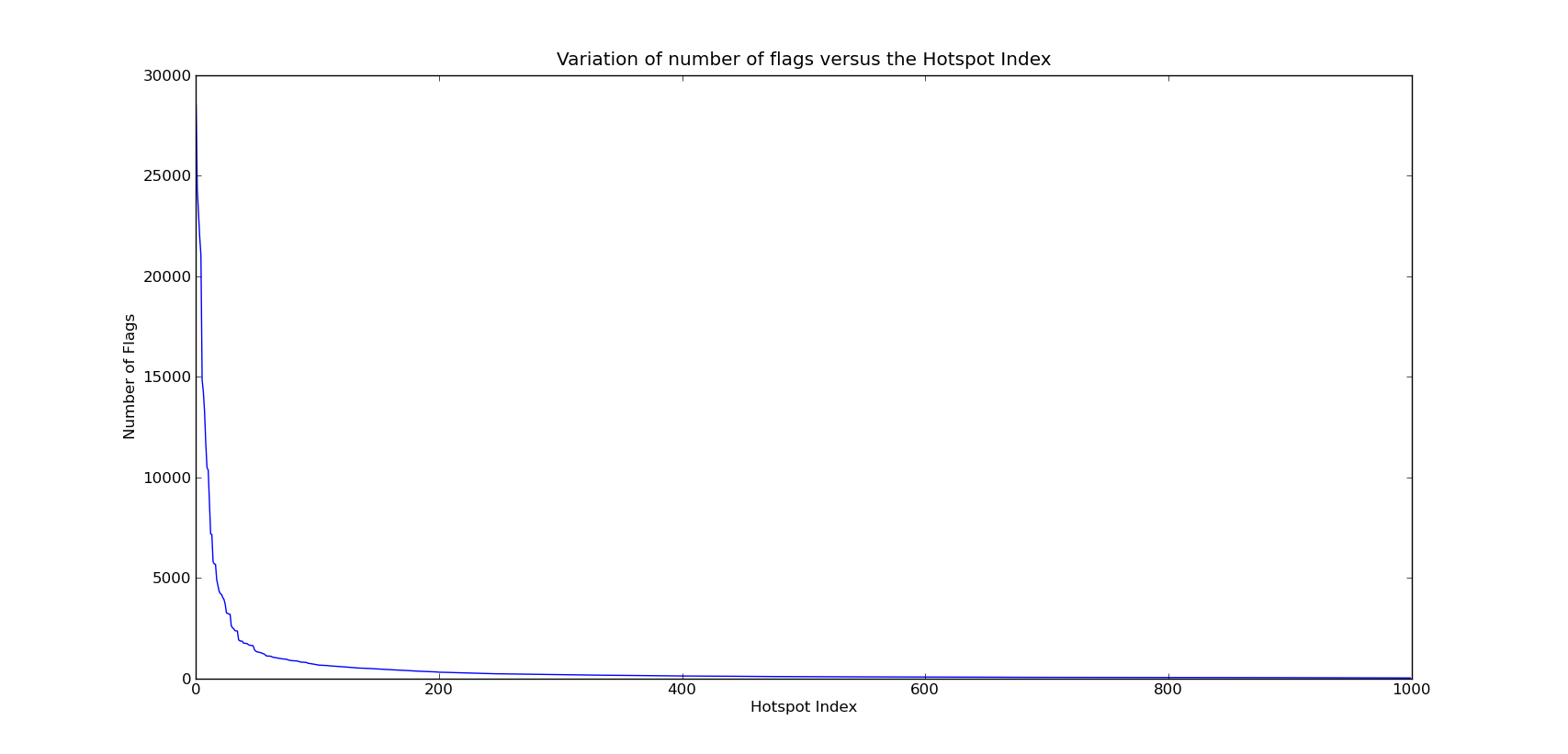}
\caption{Plot of variation of number of flags versus $HotSpotIndex$ for 
a Scale Free Graph of 1000 vertices and 4 connections.}
\label{4_optimum_ba}
\end{figure}

The $HotSpotOrdering$ list contains the vertex labels arranged in the 
decreasing order of their number of flags. Let $HotSpotIndex(u)$
denote the position of $u$ in $HotSpotOrdering$. 

Let $HotSpotOrdering = \{h_1, h_2, h_3,.... h_{|V|}\}$.\\
Let $HotSpotIndex(h_i)=i$.

The plots in Figure.~\ref{4_optimum_er} and~\ref{4_optimum_ba} represents a variation
of $flag$ values Vs $HotSpotIndex$.

Let $\alpha = |HotSpots|$. As $\alpha$ decreases, the navigation through 
the network during the navigation phase (Section.~\ref{sec:4_navigation_phase}) 
becomes increasingly difficult since it takes a longer path to reach 
vertices in $HotSpots$. As $\alpha$ increases, the difficulty involved 
in computing the all pairs shortest path between the elements of 
$HotSpots$ using Djikstra's Algorithm increases cubically, $O(\alpha^3)$. 
There is thus a need to determine an optimal value for $\alpha$.

By analyzing the plot of number of flags versus $HotSpotIndex$, we see 
that the curve takes a sharp turn at the initial values of $HotSpotIndex$ 
in the above figures.  This transition clearly shows that there are some 
vertices in the graph with relatively higher importance. 
That is, the rate of decrease of flagging decreases drastically at the 
point where the curve takes a sharp turn. Further addition of vertices 
into $HotSpots$ doesn't contribute significantly towards 
improving the efficiency of our algorithm. In fact, it causes a computational 
overhead during the creation of $HotSpotLookup$. Hence, from this point 
onwards, further addition of the vertices corresponding to $HotSpotIndex$ to 
$HotSpots$ becomes redundant. Hence, we set $\alpha$ to a value of the 
curve at which it takes a sharp turn (i.e. a drastic slope change value).

Figure.~\ref{4_optimum_er}  illustrates the flag distribution for an 
Erdos-Renyi network of 1000 vertices and an edge probability of 0.15. The plot 
shown here takes a sharp turn at the 33rd hotspot. Hence, $\alpha$ is 
set to 33. Figure.~\ref{4_optimum_ba}  illustrates the flag distribution for 
a scale-free graph of 1000 vertices and 4 connections. The plot shown here 
takes a sharp turn at the 100th hotspot. Hence, $\alpha$ is set to 100.

\subsection{Comparison Between Various Navigation Techniques}

This section highlights the effectiveness of greedy navigation over 
other methods of navigation. Consider two vertices $s,t \in V$. 
Let $s$ be source and $t$ be the target vertex. In the navigation phase 
of the algorithm, our aim is to establish a path between $s$ and 
$t$. Let $d(s,t)$ denote the length of the shortest path between $s$ and 
$t$. Here are a few methods one can adopt to accomplish the task of 
establishing a path between $s$ and $t$:

\subsubsection {1-Way Random Walk}
The idea here is to start from the source vertex $s$ and take a random walk 
$W_s$ until the target vertex $t$ is reached. Since this technique involves 
a single random walk, we refer to it as a 1-way random walk. Let 
$\beta_{s,t}$ denote the path length of the path corresponding to the random 
walk $W_s$ from $s$ to $t$. Let $\beta$ be the 
average ratio of length of 1-way random walk and length of the shortest 
path, taken over all unordered vertex pairs $(s,t)$, such that $s,t \in V$.
$$\beta = \frac{1}{\left(\begin{array}{c} |V|\\ 2\end{array}\right)} \sum_{s,t \in V} 
\frac{\beta_{s,t}}{d(s,t)}$$

\begin{figure}[htp]
\centering
\includegraphics[scale=0.14]{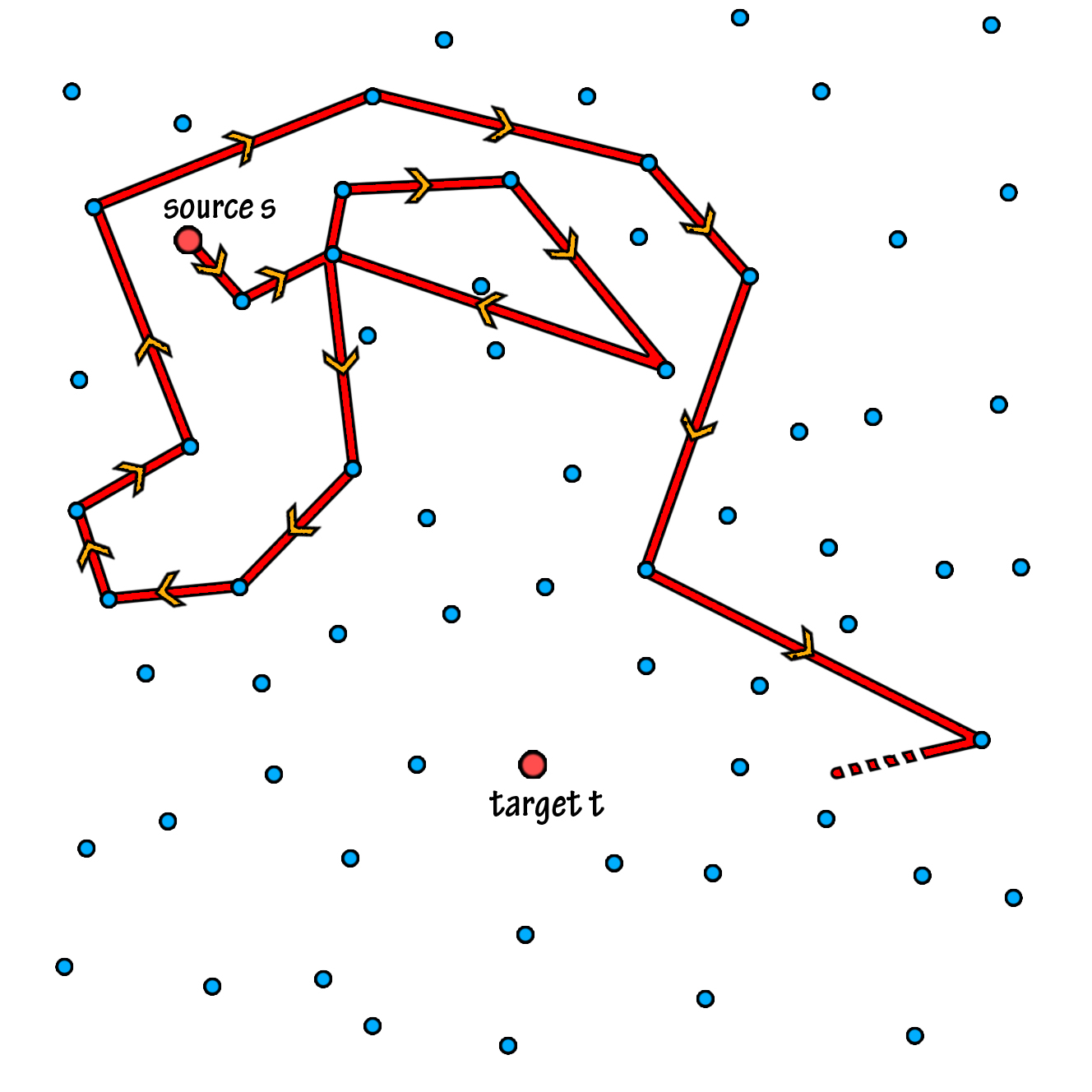}
\caption{1-way Random Walk}
\label{4_one_way}
\end{figure}

Figure.~\ref{4_one_way} illustrates the technique of 1-way random walk from $s$ to $t$. $W_s$
terminates when it reaches $t$. $W_s$ is indicated by the red walk. The source $s$ 
and target $t$ is denoted by the red dots. The blue dots indicate the other vertices 
in the network.

\subsubsection{2-Way Random Walk}

After taking random walks $W_s$ and $W_t$ from $s$ and $t$ simultaneously 
until the two walks intersect, one can find a path from $s$ to $t$.

This idea is similar to the one implemented in the learning phase in 
Section ~\ref{sec:4_learning_phase}. Let $\gamma_{s,t}$ denote the length of the path thus obtained. 
Let $\gamma$ be the average ratio of length of 2-way Random Walk and the 
length of the shortest path taken over all the unordered vertex pairs $(s,t)$, such that $s,t \in V$.

$$\gamma = \frac{1}{\left(\begin{array}{c} |V|\\ 2\end{array}\right)} \sum_{s,t \in V} \frac{\gamma_{s,t}}{d(s,t)}$$

\begin{figure}[htp]
\centering
\includegraphics[scale=0.15]{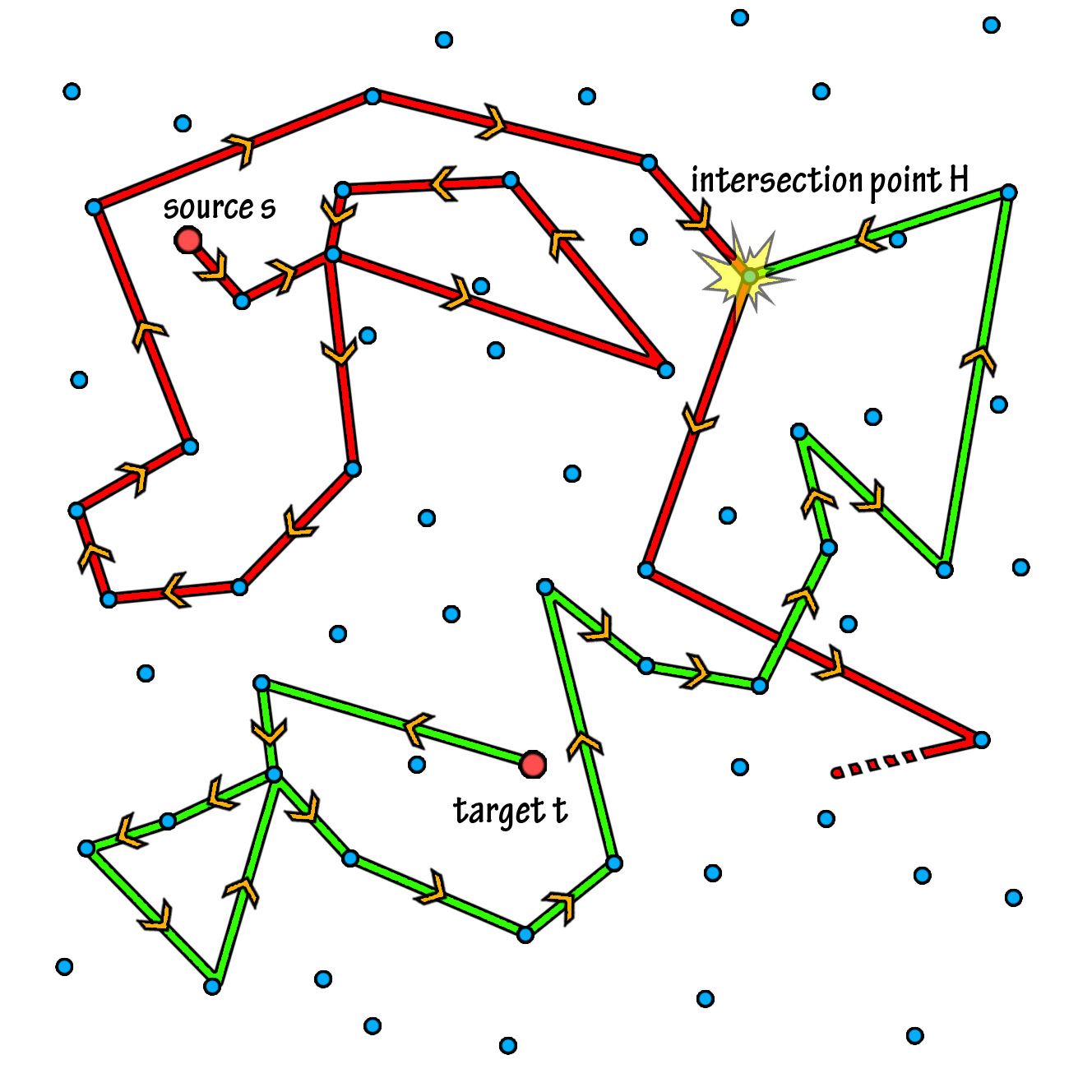}
\caption{2-Way Random Walk}
\label{4_two_way}
\end{figure}

Figure~\ref{4_two_way} illustrates the technique of 2-way random walk between $s$ and $t$. 
$W_s$ and $W_t$ are constructed simultaneously until they intersect.
The intersection point of the two walks is indicated by $H$. $W_s$ and $W_t$ 
are indicated by the red walk and green walk respectively. Source $s$ and the 
target $t$ are denoted by the red dots. 

\subsection{Comparison between PCA and Degree Based Navigation (Lada A. Adamic)}
Degree based navigation assumes that we have only local knowledge of the network. 
To pass a message from a source to a target, the source node passes the message to the neighbor with the highest degree. 
This process continues till the target is found. 
However, the message may not be passed to a node already visited, if it has other available neighbors. 
If there are no available neighbors, the message may be passed to a node that is already visited. 

\begin{figure}[htp]
\centering
\includegraphics[scale=0.4]{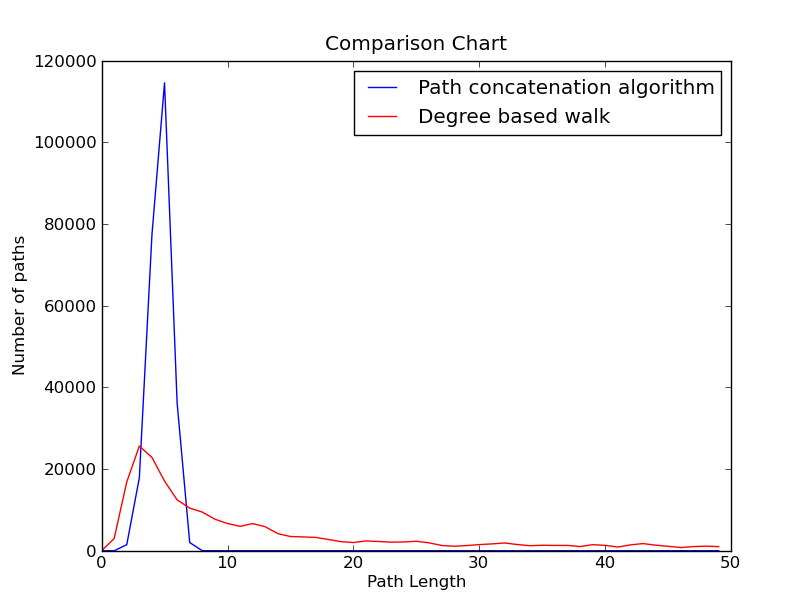}
\caption{Comparison between PCA and Degree Based Navigation}
\label{pca_vs_adamics}
\end{figure}

We compare the path lengths obtained from this type of navigation to the path lengths obtained using PCA for a Scale-Free network of 500 nodes, considering all the vertex pairs. Figure~\ref{pca_vs_adamics} illustrates the empirical results that we have obtained.The average path length for PCA and degree based navigation was found to be 4 and 24 respectively.

As evident from the plot, degree based navigation yields paths of much greater lengths than the PCA paths. 
This is due to the fact that the degree based approach spends a lot of hops between nodes of high degree after a brief initial period of exploring fresh nodes. 

\subsubsection{Path Concatenation Algorithm}
Let $\delta_{s,t}$ denote the length of the path from $s$ to $t$ obtained after 
executing the path concatenation algorithm. Let $\delta$ be the average ratio of 
length of the path obtained by the path concatenation algorithm and the length 
of the shortest path, taken over all the unordered vertex pairs $(s,t)$, such 
that $s,t \in V$.

$$\delta = \frac{1}{\left(\begin{array}{c} |V|\\ 2\end{array}\right)} 
\sum_{s,t \in V(G)} \frac{\delta_{s,t}}{d(s,t)}$$

\begin{figure}[htp!]
\centering
\includegraphics[scale=0.2]{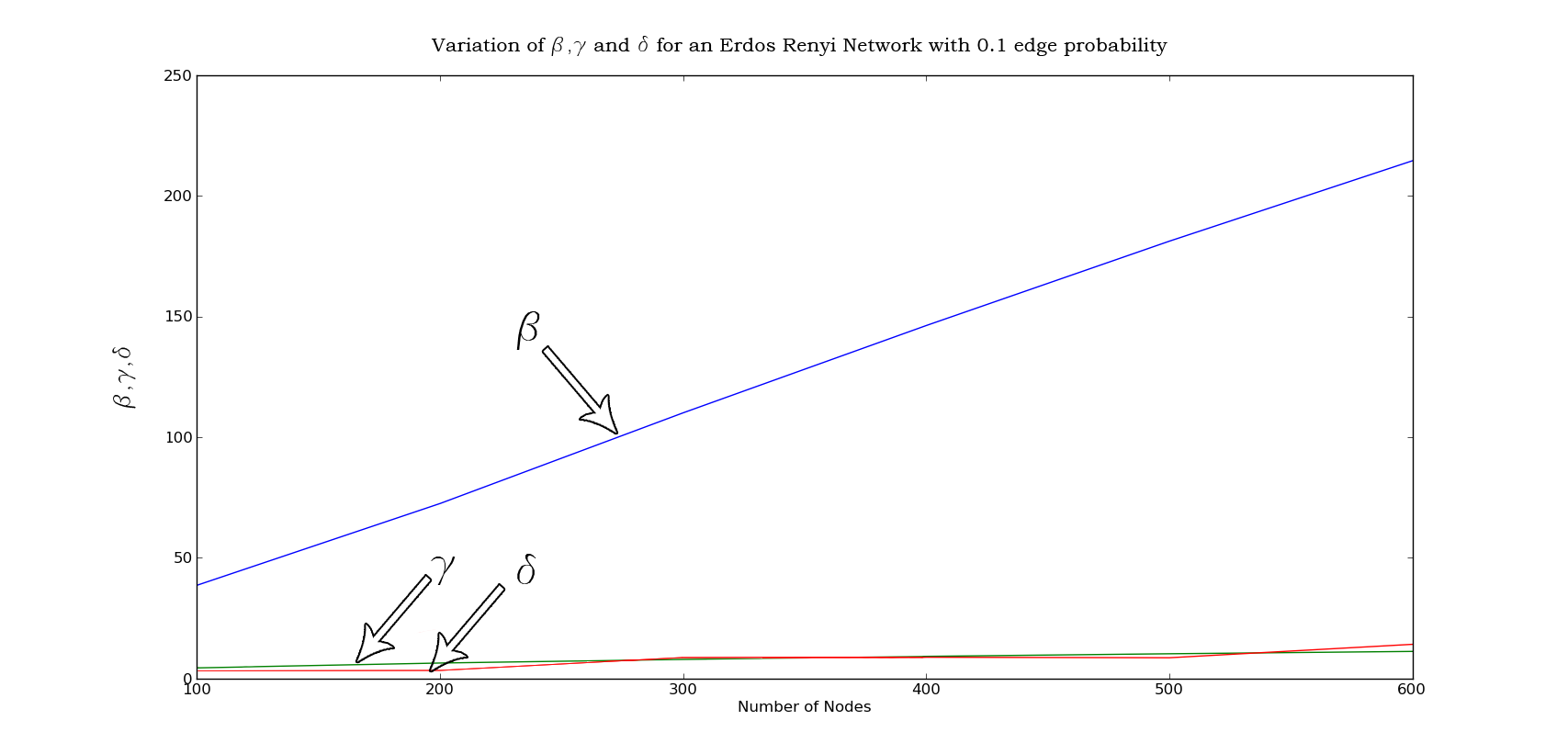}
\caption{Plot of $\beta$, $\gamma$ and $\delta$ 
versus the number of vertices for an Erdos-Renyi network with an edge probability of 0.1}
\label{4_performance_er}
\end{figure}

\begin{figure}[htp!]
\centering
\includegraphics[scale=0.2]{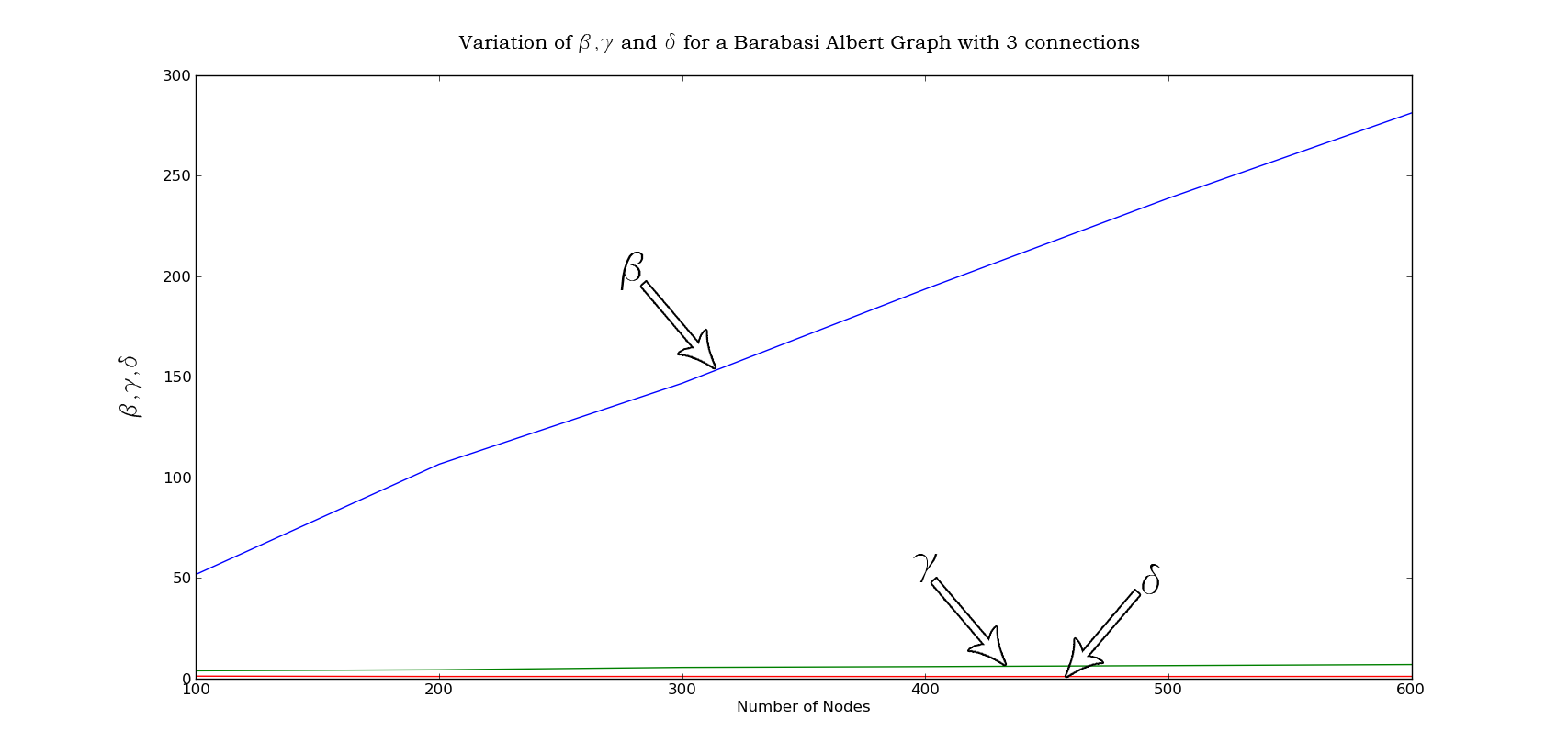}
\caption{Plot of $\beta$, $\gamma$ and $\delta$ versus the number
of vertices for a Barabasi-Albert Graph with 3 connections}
\label{4_performance_ba}
\end{figure}

\begin{figure}[htp!]
\centering
\includegraphics[width=.23\textheight]{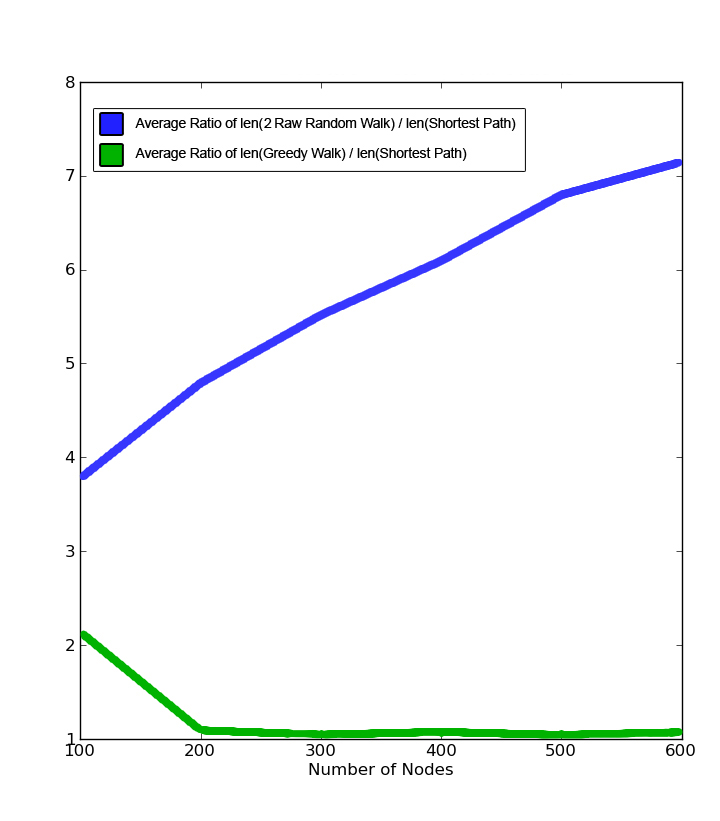}
\caption{This plot is a zoomed version of previous plot in 
Figure.~\ref{4_performance_ba} of  $\gamma$ and $\delta$ versus the number
of vertices for a Scale-free Graph with 4 connections}
\label{4_performance_ba_zoom}
\end{figure}

Figure.~\ref{4_performance_er} is a plot of variation of $\beta$, 
$\gamma$ and $\delta$ versus the number of vertices for an Erdos-Renyi network 
with an edge probability of 0.1. Figure.~\ref{4_performance_ba} is a plot of variation of 
$\beta$, $\gamma$ and $\delta$ versus the number of vertices for a 
scale free network with 4 connections. The blue curve indicates $\beta$, green curve 
indicates $\gamma$ and the red curve indicates $\delta$. From the plots, it is clear 
that the $\delta$-curve always lies below the $\beta$ curve. This implies that, the 
greedy navigation performs better than 1-way Random Walk.

In case of Erdos-Renyi networks, the $\delta$ curve and the $\gamma$ curve lie very 
close to each other, and also intersect at certain points in the plot. Hence, the 
performance of the proposed algorithm is as good as a 2-way random walk technique.

In case of scale-free graphs, the $\delta$ curve always lies below that 
of the $\gamma$ curve. Hence, the performance of the greedy technique is better than 
that of the 2-way random walk technique.

\subsection{Hotspot Distribution for Erdos-Renyi and Scale-free graphs}

The hotspot set we introduced exhibit some characteristic features. We note that 
the proposed path concatenation algorithm doesn't perform significantly 
better than the 2-way random walk. The reason being that the Erdos-Renyi graphs do not 
contain vertices of comparatively higher degree than the rest of the vertices in the graph. 
Whereas in case of a scale-free graph, some vertices have very high degree (\emph{hubs})
while the rest have comparatively low degree. These \emph{hubs} form valid candidates for 
the hotspot set. They can be reached in a few hops from any vertex in the graph, 
thus making the PCA an efficient strategy.

\begin{figure}[htp]
 
\begin{center}
\includegraphics[scale=0.22]{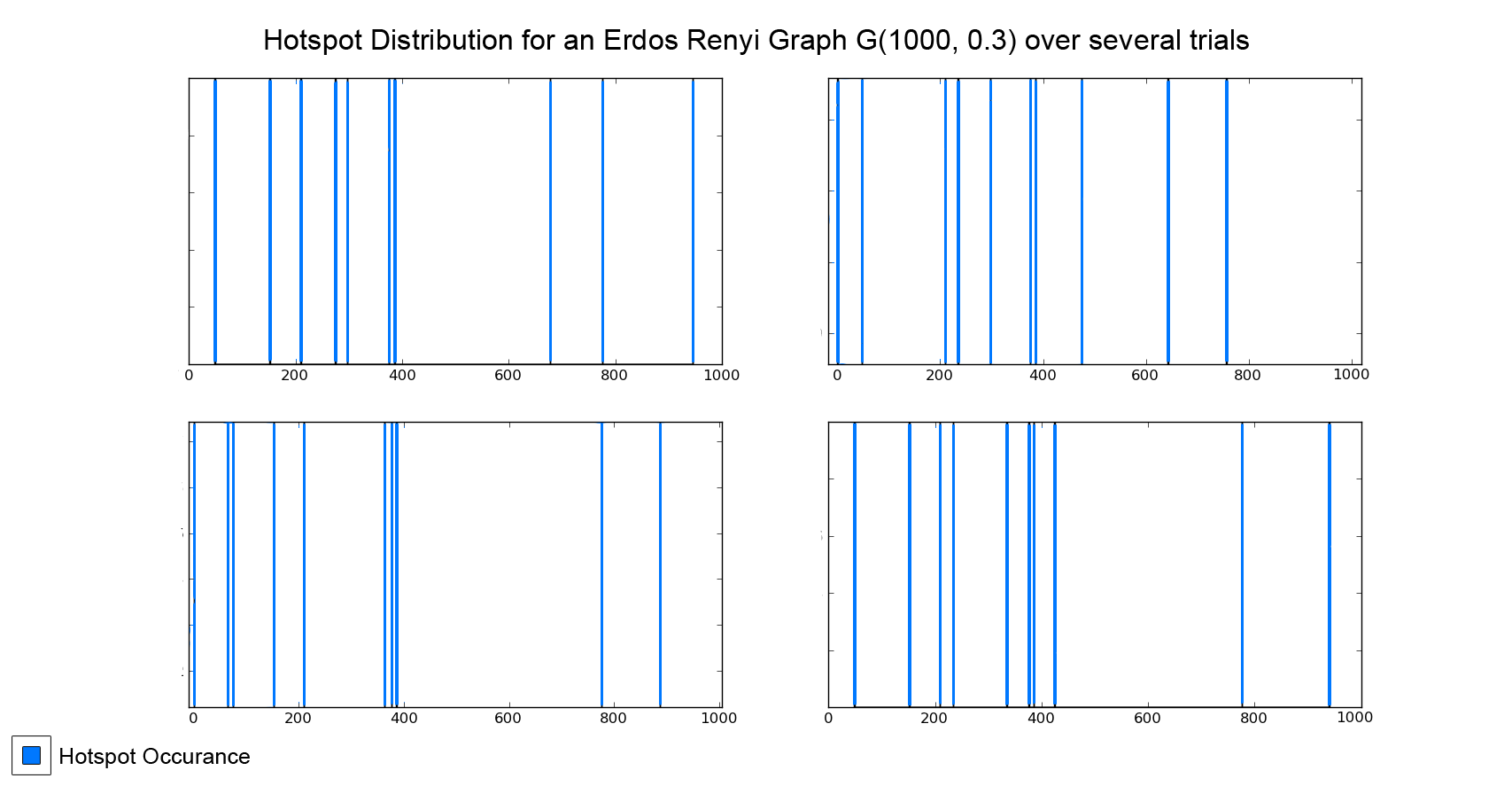}
\caption{Hotspot Distribution for an Erdos-Renyi Graph}
\label{4_hotspot_er}
\end{center}

\end{figure}

\begin{figure}[htp]
\begin{center}
\includegraphics[scale=0.22]{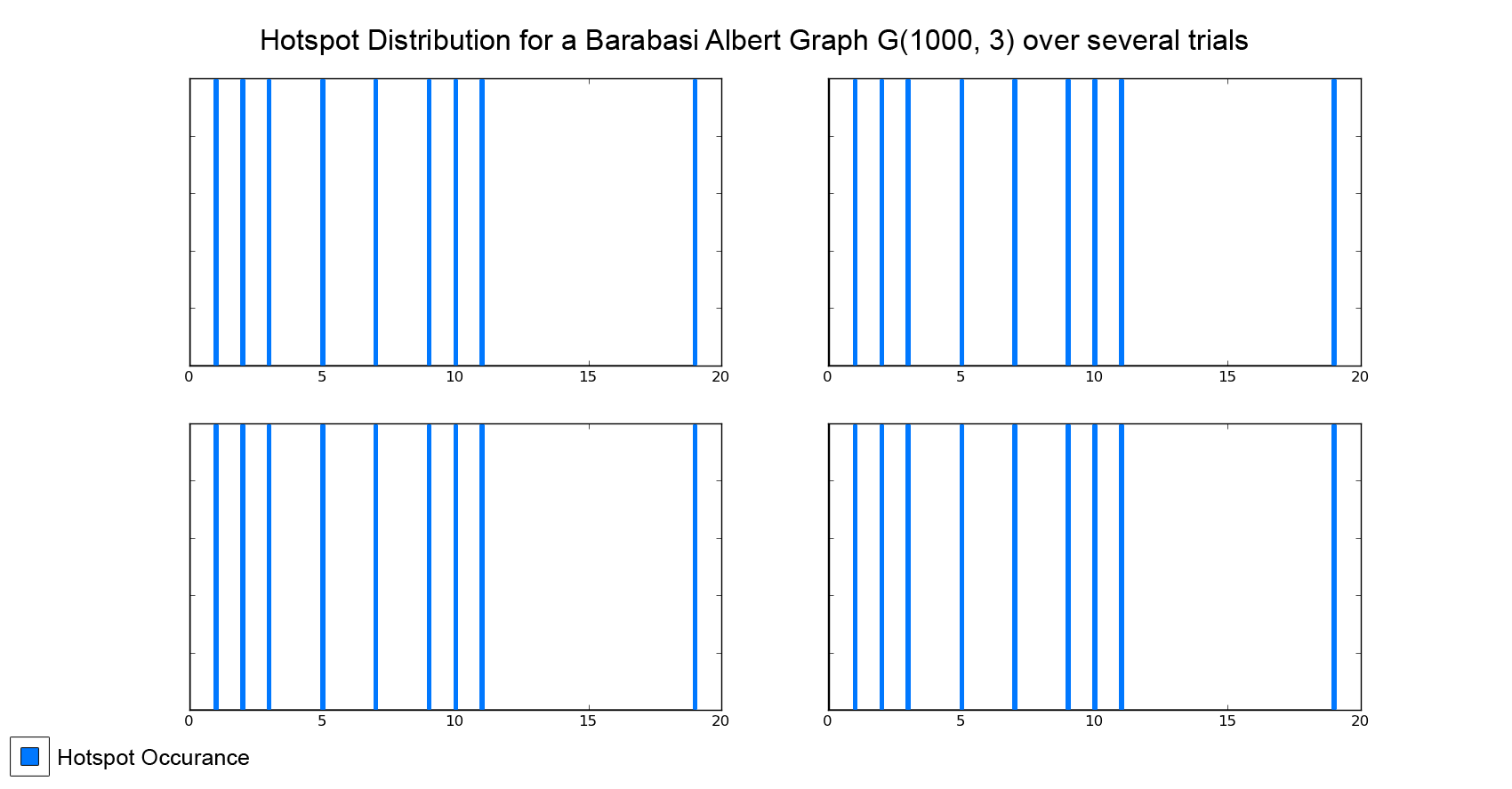}\\
\caption{Hotspot Distribution for a Scale-free Graph}
\label{4_hotspot_ba}
\end{center}
\end{figure}

Fixing an Erdos-Renyi graph $G$ on 1000 vertices with probability 
$p=0.3$, as we run the learning phase of the PCA algorithm 4 times on the same graph, 
we note that the hotspot sets are different each time we run the algorithm. 
The top hotspot set is shown in Figure.~\ref{4_hotspot_er}, the x-axis denotes the vertex name 
and the vertical lines represent the hotspot. E.g. if a vertex, say 306, 
has a vertical line, it means that the vertex is chosen as a hotspot. 

Repeating the same procedure on a scale-free graph, we note that the top hotspot set
remains more or less the same as shown in Figure.~\ref{4_hotspot_ba}

\section{Center-strategic Paths on Scale Free Networks}

Given a path $(v_1,v_2,v_3,...,v_k)$ with closeness centrality 
values $(c_C(v_1),c_C(v_2),...,c_C(v_k))$ as defined in Section. \ref{sec:4_notions},
we compute the ranking of vertices $(R(v_1),R(v_2),R(v_3),...,R(v_k))$. By 
\emph{rank-plot} of this path, 
we mean a plot of the ranks $(R(v_1),R(v_2),R(v_3),...,R(v_k))$. 
If the rank-plot of a path has no more than one maxima, we call such a path 
a \emph{center-strategic path}. 

The algorithm presented in this article is inspired by the strategy
adapted by humans to learn the centers of a network and navigate in a 
center-strategic way. The presented algorithm resonates with the technique used 
by humans. We establish this fact by showing that the PCA algorithm yields 
center-strategic paths on scale free networks.

\begin{figure}[htp!]
 
\begin{center}
\includegraphics[width=.4\textheight]{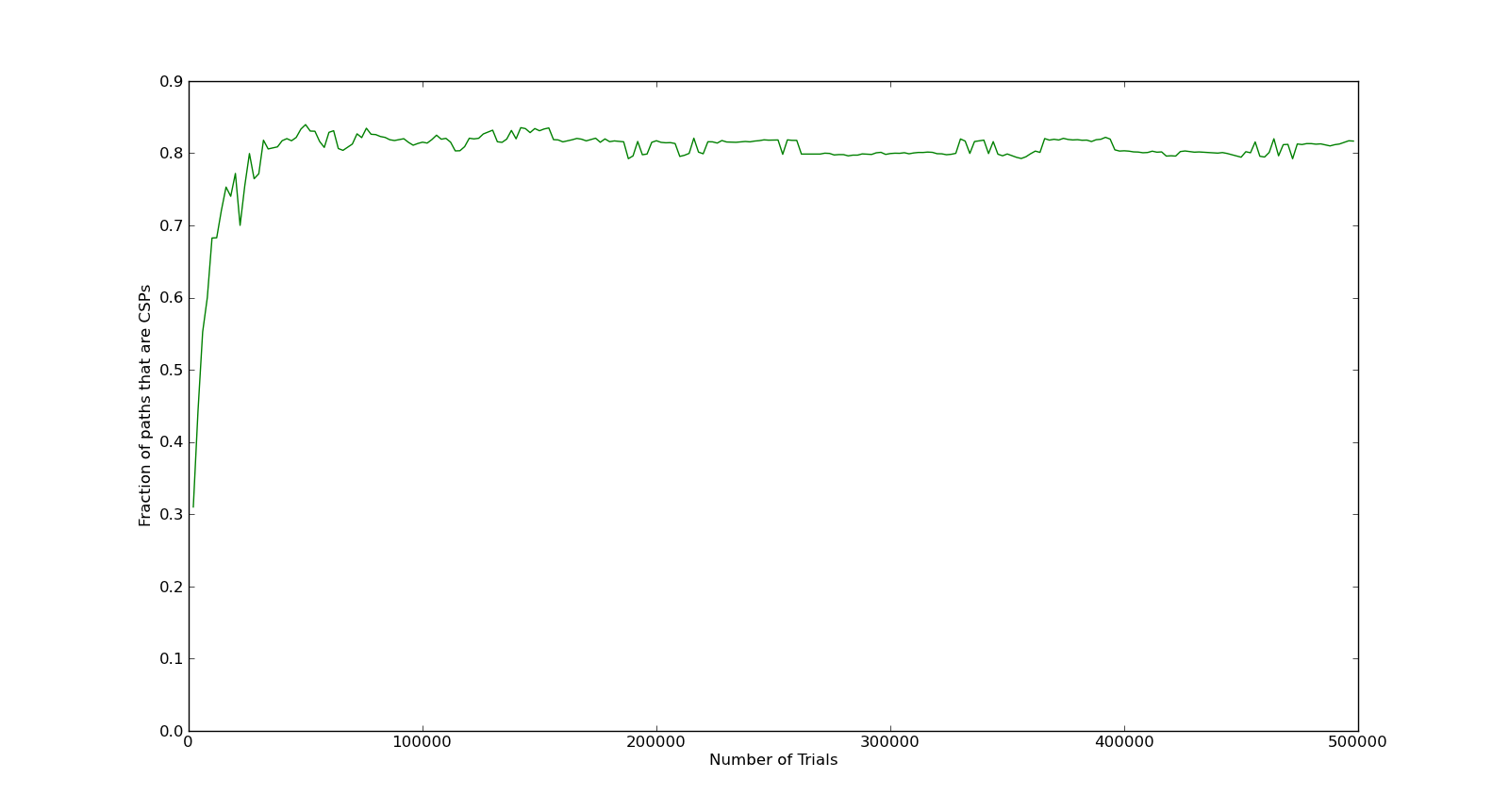}
\caption{Center Strategic Paths Using Path Concatenation Algorithm}
\label{4_mainplot}
\end{center}
\end{figure}

Consider a scale-free graph $G(V,E)$. Let $k$ denote the number of iterations 
(or trials) during the navigation phase of the path concatenation algorithm.
After $k=100$, for every $k$, we jump to navigation phase, take $^{|V|}C_2$ greedy 
traversals (as explained in Section.~\ref{sec:4_navigation_phase}) between all vertex pairs and check for 
the number of paths that are center-strategic. The ratio of the number of 
center-strategic paths to the number of greedy traversals $^{|V|}C_2$ is denoted 
by $\psi$.

Figure.~\ref{4_mainplot} shows the variation of $\psi$ as k increases. We note that the path 
concatenation algorithm yields center-strategic paths 80\% of the times on 
scale free networks.

\section{Conclusion}
\label{sec:4_conclusion}

Motivated by the strategy adopted by humans to navigate in an unknown environment, we presented
in this paper, an algorithm which simulates human navigation. We showed that the algorithm
performs better than the 1-way and 2-way random walk technique. We further showed that 
the proposed algorithm generates center-strategic paths on scale-free networks. 

A possible further work would be to study the convergence of the learning phase of 
the algorithm. One can start with the basic graph structures like the paths, grids, trees 
and study the hotspot distribution. It would also be interesting to classify the networks
on which the algorithm performs well and those on which it does not. One can study the 
performance of PCA algorithm on real - world networks as compared to other other 
navigational techniques. 

\bibliographystyle{plain}  
\bibliography{references}

\end{document}